# Performance of MF-MSK Systems with Pre-distortion Schemes

**Labib Francis Gergis**

Misr Academy for Engineering and Technology, Mansoura, Egypt

drlabeeb@yahoo.com

*Abstract:*

Efficient RF power amplifiers used in third generation systems require linearization in order to reduce adjacent channel inter-modulation distortion, without sacrificing efficiency. Digital baseband predistortion is a highly cost-effective way to linearize power amplifiers (PAs).

New communications services have created a demand for highly linear high power amplifiers (HPA's). Traveling Wave Tubes Amplifiers ( TWTA ) continue to offer the best microwave HPA performance in terms of power efficiency, size and cost, but lag behind Solid State Power Amplifiers ( SSAP's ) in linearity. This paper presents a technique for improving TWTA linearity. The use of pre-distorter (PD) linearization technique is described to provide TWTA performance comparable or superior to conventional SSPA's. The characteristics of the PD scheme is derived based on the extension of Saleh's model for HPA. The analysis results of Multi-frequency Minimum Shift Keying (MF-MSK) in non-linear channels are presented in this paper.

## 1. Introduction

Power amplifiers are essential components in communication systems and are inherently nonlinear. The nonlinearity creates spectral growth beyond the signal bandwidth, which interferes with adjacent channels. It also causes
distortions within the signal bandwidth, which decreases the bit error rate at the receiver. In digital predistortion system, an inverse characteristic of power amplifier is generated and its amplitude and phase are combined to the signal input. So the input signal is predistorted and the power amplifier response is corrected. This process has to be controlled at high accuracy to achieve a satisfactory compensation effect. The inverse characteristics are stored in a memory (look-up table) and this data are updated using an error that is produced by comparing the outputs of power amplifier with the input signals.

Power amplifiers (PA's) are vital components in many communication system. The linearity of a PA response constitutes an important factor that ensures signal integrity and reliable performance of the communication system. High power amplifiers (HPA) suffer from the effects of amplitude modulation to amplitude modulation distortion (AM/AM), and amplitude modulation to phase modulation distortion (AM/PM) [1], and [2] during conversions caused by the HPA amplifiers. These distortions can cause intermodulation (IM) components, which is undesirable to system designs. The effects of AM/AM and AM/PM distortions can cause the signal distortion that degrade the bit error rate performance of a communication channel.

The modulation schemes of wireless communication systems such as GSM are mainly derived from the class of continuous phase modulated ( CPM ) signals with modulation index $h$

= 0.5, which have excellent properties in non linear channels and are power efficient, but their bandwidth efficiency is rather low[10]and[11]. The bandwidth efficiency can be increased either by increasing the number of signal phase levels, or by increasing the number of signal amplitude levels. The latter has the drawback that the signal envelope is not constant, and therefore non linear amplification may cause spreading of the signal spectrum and additional increase in system bit error rate. A modulation scheme that produces a constant envelope continuous phase signal set, can be implemented by quadrature-carrier multiplexing of two frequency/phase modulated signal of the type *NFSK / 2PSK*, both with the same frequency in each transmission interval. The generated signal can be viewed in each transmission interval as MSK signal at one of the *N* frequencies and is referred to as Multi-Frequency Minimum Shift Keying ( *MF-MSK* )[3], which is combining a number of attractive attributes such as constant envelope, excellent spectral properties, high power efficiency, and self synchronization capability.

The amplitude and phase modulation distortions are minimized using linearization method. The linearization method requires modeling the characteristics of the amplitude distortion and phase distortion of the HPA. A Saleh model has been used to provide the linearization method and applied to measured data from HPA that characterize the distortion caused by the HPA. The measured data provides a performance curve indicating nonlinear distortion. The forward Saleh model is a mathematical equation that describes the amplitude and phase modulation distortions of the HPA [4], and[12-13]. The amount of desired linearization is then determined to inversely match the amount of distortion for canceling out the distortion of the HPA. This paper investigates Pre – distorter technique for the linearization of a HPA to mitigate the AM/AM and AM/PM effects in digital communication systems.

The remainder of this paper is organized as follows. In section 2, a description of the proposed system was discussed. In section 3, the non-linear model for a HPA, and pre-distortion scheme were driven. Performance analysis and numerical results were presented in sections 4, and 5. Section 6, showed the improvement in the performance of MF-MSK transmission system over the nonlinear channels.

## 2. System Model

Fig. 1 illustrates the baseband equivalent functional block diagram of the MF-MSK transmitted signal through HPA using pre-distorter scheme.

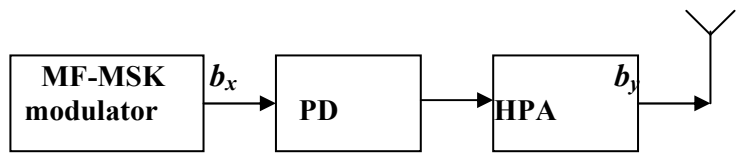

Fig. 1. Simplified MF-MSK Transmitter through HPA

Let the base-band input signal to HPA could be modeled as

$$\left| b_x(t) \right| = U_x(t)\, e^{j\alpha_x(t)} \tag{1}$$

where $U_x(t) = b_x(t)$.

$$b(t, \alpha) = (\sqrt{2E/T}) \cos[2\pi f_c t + \Phi(t, \alpha)] \tag{2}$$

whish has a constant envelope $\sqrt{2E/T}$, where $E$ is the average signal energy per symbol interval, $T$ is the length of symbol interval, $f_c$ is the carrier frequency. $\alpha$ is the transmitted *M*-ary data symbol sequence, with symbols taken from the set $\alpha_i$
$\in \{\pm 1, \pm 3, \pm 5, ....., \pm(M-1)\}$, $h$ is modulation index = 0.5. For a symbol interval *n*, where t $\in$ [*NT, (n+1) T*]. The information is carried by the phase of the signal

$$\Phi(t, \alpha) = 2\pi \sum_{i=-\infty}^{\infty} \alpha_i h_i q(t - iT) + \theta_n \tag{3}$$

*q(t)* is phase response equals

$$q(t) = \begin{cases} 0 & t \leq 0 \\ 1/2 & t > T \end{cases} \tag{4}$$

## 3. Nonlinearity Effects on MF-MSK Signal

The classical and most often used nonlinear model of power amplifier is Saleh's model [3]. It is a pure nonlinear model without memory. The equations define this base-band model of HPA as two modulus dependent *transfer functions* are defined as[6] :

$$A[U_x] = \alpha_a U_x / 1 + \beta_a U_x^2$$
$$\Phi[U_x] = \alpha_\Phi U_x / 1 + \beta_\Phi U_x^2 \tag{5}$$

where $A[U_x]$ and $\Phi[U_x]$ are the corresponding AM/AM and AM/PM characteristics respectively, both dependent exclusively on $U_x$, which is the input modulus to HPA.
The values of $\alpha_a$, $\beta_a$, $\alpha_\Phi$ and $\beta_\Phi$ are defined in [7]. The corresponding AM/AM and AM/PM curves so scaled are depicted in Fig. 2.

The HPA operation in the region of its nonlinear characteristic [12], causes a nonlinear distortion of a transmitted signal, that subsequently results in increasing the bit error rate (*BER*), and the out-of-band energy radiation ( spectral spreading ).

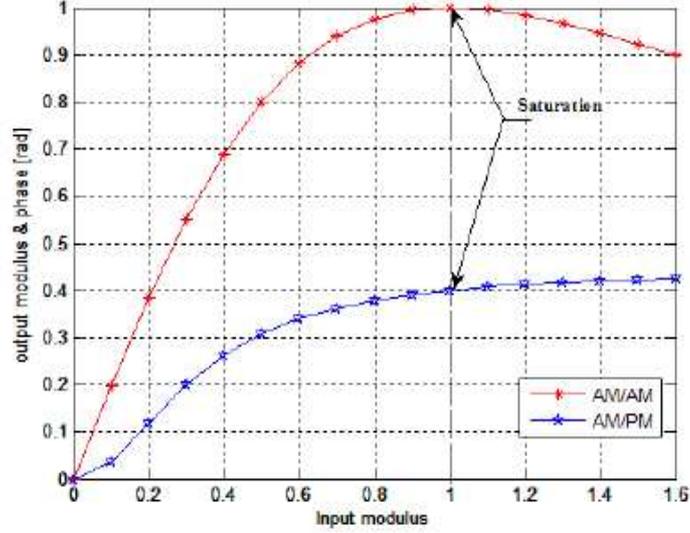

Fig. 2. AM/AM and AM/PM normalized charateristics of the Saleh model
For TWTA HPA's

The operating point of HPA is defined by input back-off (*IBO*) parameter which corresponds to the ratio of of saturated input power ($P_{max}$), and the average input power ( $P_{in}$ ) [6] :

$$IBO_{dB} = 10 \log_{10} \ (P_{max} / \ P_{in}) \qquad (6)$$

The measure of effects due to the nonlinear HPA could be decreased by the selection of relatively high values of *IBO*

The output of HPA defined in Fig. 1, is expressed as

$$b_y = A \ [\ U_x\ ] \ e^{j(\alpha x + \Phi[U_x])} \qquad (7)$$

where the input-output functional relation of the HPA has been defined as a *transfer function*. Hence in order to obtain linearization, it may be necessary to estimate a discrete inverse multiplicative function HPA$^{-1}$ [.] such that

$$b_x = b_y \ . \ \text{HPA}^{-1} \ [U_y] \qquad (8)$$

An alternative expression for the AM/AM distortion in (5), convenient for the theoretical formulation of the linearizer, is obtained by replacing the saturation input amplitude $As = 1 / \sqrt{\beta_s}$ in the expression (8). This gives

$$A[U_x] = A^2_s \ \alpha_a \ U_x \ / \ A^2_s + \ U^2_x \qquad (9)$$

The theoretical AM/AM inverse transfer function $A^{-1}$ [.] could be determined by solving (9) for $U_x = A \ \{\ A^{-1} \ [U_x]\ \}$

$$A^{-1}[u] = (A_s^2 \alpha_a / 2U) \left[ 1 - \sqrt{1 - (2U / A_s \alpha_a)^2} \right] \quad (10)$$

Considering the alternative configurations shown in Fig. 3, where the same input-output function is applied as a pre-distorter [PD] for the linearization of the same HPA. Letting ψ[.] denote the AM/PM characteristic of the PD block.
For the case of a Pre-distortion, we have[7] :

$$b_{pout} = A^{-1}[U_x] e^{j(\alpha x + \psi[U_x])} \quad (11)$$

$$b_y = A\left[ A^{-1}[U_x] \right] e^{j(\alpha x + \psi[U_x] + \Phi[A^{-1}[U_x]])} \quad (12)$$

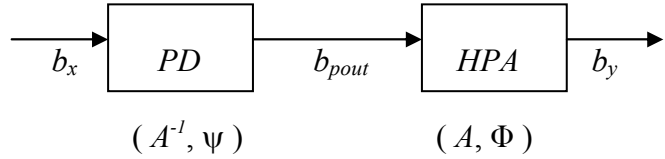

Fig. 3. Pre-distortion For HPA Linearization

The ideal AM/PM correction requires that

$$\psi[U_x] = -\Phi\{ A^{-1}[U_x] \} \quad (13)$$

$$b_{pin} = A[U_x] e^{j(\alpha x + \Phi[U_x])} \quad (14)$$

$$b_y = A^{-1}\{ A[U_x] \} e^{j(\alpha x + \Phi[U_x] + \psi[A^{-1}[U_x]])} \quad (15)$$

Pre-distortion linearization idea, as depicted in Fig. 4, can be used to linearize over a wide bandwidth. This is achieved by pre-distortion of the signal prior to amplification with the inverse characteristics of the distortion that will be imposed by the power amplifier. Thus the output of the HPA is a linear function of the input to the predistorter [8],[10], and [13].

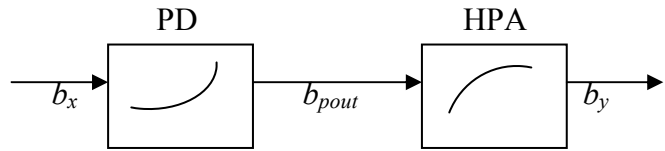

Fig. 4. Basic System Functional Diagram of Pre-distortion Linearization

where the AM/PM correction for Post-distorter case, requires that

$$\psi[U] = -\Phi\{ A^{-1}[U] \} \quad (16)$$

A description of the ideal theoretic AM/AM and AM/PM inverse characteristics, valid for the normalized Saleh's HPA model is shown in Fig. 5

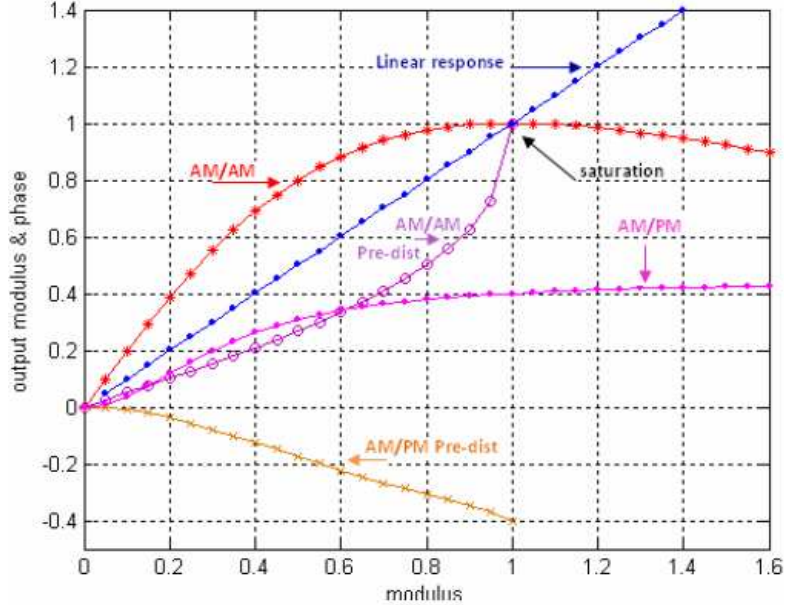

Fig. 5. AM/AM and AM/PM pre-distortion for the Saleh model

## 4. Bit Error Rate Analysis

The error probabilities of MF-MSK and *NFSK / 2PSK* are the same if the signal set sizes *M = 4N*. For a MF-MSK system, an upper bound on the symbol error probability is given [3] as

$$P_{es} \leq (M-2) \, Q\left(\sqrt{d^2_{min} E_s / N_o}\right) \qquad (17)$$

where $E_s = E_b \log_2 M = E_b (2 + \log_2 N)$ is the symbol energy, $E_b$ is the energy per bit, $N_o$ is the one-sided power spectral density of the input AWGN, and the *Q* function can be represented as

$$Q(x) \leq (1/2) \, e^{-x/2}$$

Consequently, for the bit error probability we **have**

$$P_{eb} \leq (2N-1) \, Q\left(\sqrt{d^2_{min}(E_b/N_o)(2+\log_2 N)}\right) \qquad (18)$$

## 5. Numerical Results

In this section, we have proposed and analyzed the PD linearizer which compensates for an arbitrary, invertible, channel nonlinearity in radio systems employing MF-MSK. Fig. 6, illustrates the performance analysis for MF-MSK, it is shown the improvement in the bit error rate through the increasing in M values. A performance comparison for 16QAM scheme, 16PSK scheme[9], and varieties of MF-MSK (16, 32, and 64MSK schemes) are reported in Fig. 7. It is clear to demonstrate the superiority type of MF-MSK analysis over other types of digital modulation schemes. The performance of the MF-MSK transmission system expressed by bit error probability ($P_e$) versus $E_b / N_o$ under the case of nonlinearity for different values of *IBO*

($IBO$ = 5 dB, $IBO$ = 7 dB, and $IBO$ = 9 dB), compared with the case of using pre-distorter, is illustrated by Fig. 8. It is shown that it is highly recommended to use PD at the transmitter side in order to suppress the undesirable nonlinearity effects and to get improved bit error performance.

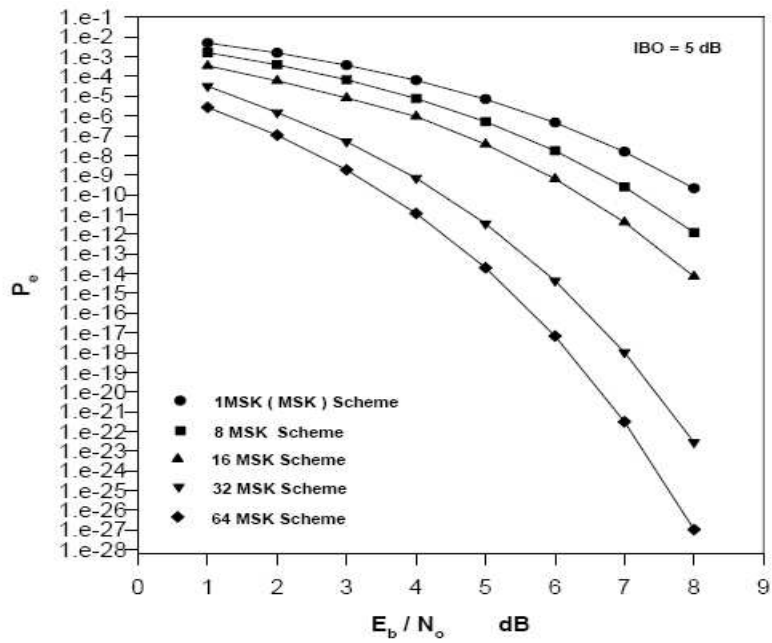

Fig. 6. Performance Analysis for MF-MSK with different M values

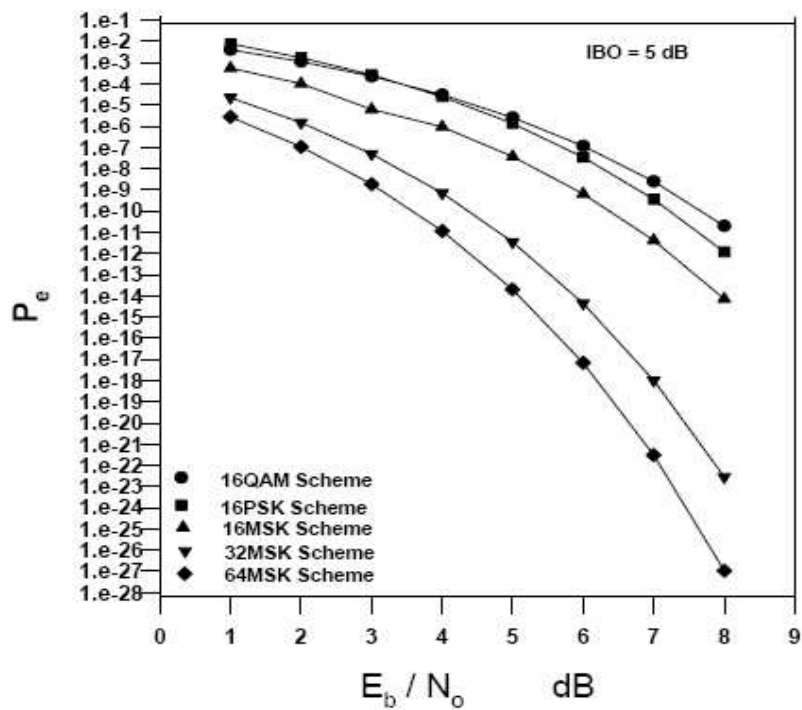

Fig. 7. Performance Comparison for different Schemes
of M-ary Digital Modulation Techniques

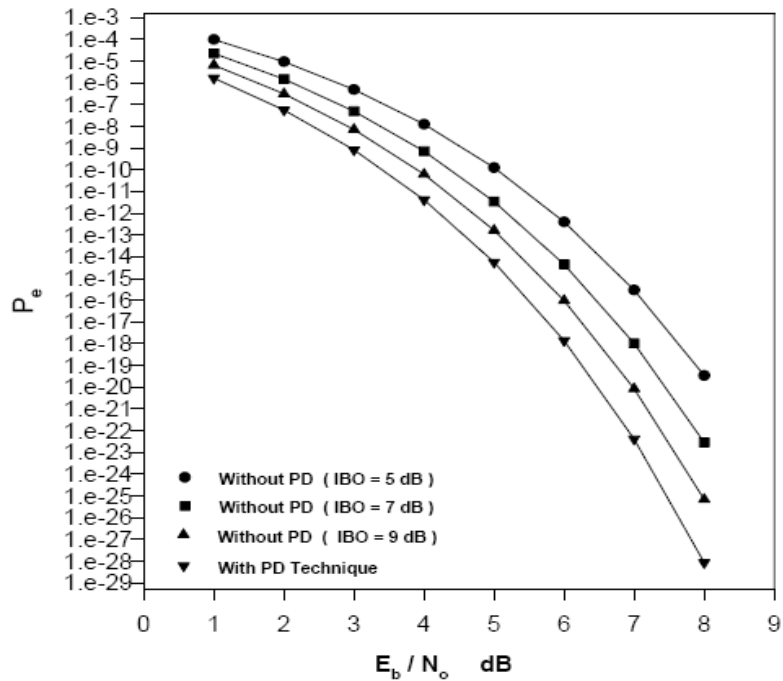

Fig. 8. Performance Analysis for MF-MSK
within different Values of IBO and with Pre-distortion Technique

## 5. Conclusions

In this paper, a spectral efficient modulation scheme has been derived; the MF-FSK signal has constant envelope and a continuous phase. It has been shown that its error performance improves as the size M of the signal set increases. With respect to MQAM and MPSK with the same signal set size M, MF-MSK always exhibits very significant gains in power. Due to large envelope variations, the distortion introduced by nonlinear HPA is more obvious in modulated systems. We presented so a useful pre-distorter design for non linear HPA's with memory that the out-of-band pre-distortion effect is very close to the ideal case. The effects of nonlinearities (AM/AM and AM/PM) were analyzed. It is shown that these effects can be compensated by using PD technique. From analytical results, it is confirmed that PD system with MF-MSK gives a good performance improvements compared to MF-MSK signals without PD.

## References

[1] P. Varahram, S. Mohammdy, M. Hamidon, R. Sidek, and S. Khatun, " An improvement Method for Reducing Power Amplifiers Memory Effects Based on Complex Gain Predistortion ", Australian Journal of Basic and Applied Sciences, VOL. 4, No. 7, pp. 2059-2067, 2010

[2] S. Chang, "An efficient compensation of TWTA's nonlinear distortion in wideband OFDM systems," IEICE Electronics Express, Vol. 6, No. 2, P 111-116, 2009.

[3] S. Fleisher, and S. Qu," Multifrequency Minimum Shift Keying, "IEEE JOURNAL ON SELECTED AREAS IN COMMUNICATIONS, VOL. 10, NO. 8, OCTOBER 1992.

[4] P. Drotar, J. Gazda, D. Kocur, and P. Galajda, "Effects of Spreading Sequences on the Performance of MC-CDMA System with Nonlinear Models of HPA, "RADIOENGINEERING, VOL. 18, NO. 1,

APRIL 2009.

[5] M. Chen, and O. Collins, " Multilevel Coding for Nonlinear ISI Channels,"IEEE TRANSACTIONS ON INFORMATION THEORY, VOL. 55, NO. 5, MAY 2009

[6] R. Zayani, and R. Bouallegue,"Predistortion for the compensation of HPA Nonlinearity with neural networks: Application to satellite communications," IJCSNS International Journal of Computer Science and Network Security, VOL. 7, No. 3, March 2007.

[7] P. Sardrood, G. solat, and P. Parvand, "Pre-distortion Linearization for 64-QAM Modulation in Ka-Band Satellite Link, "IJCSNS International Journal of Computer Science and Network Security, VOL. 7, No. 3, March 2007

[8] D. Giesbers, S. Mann, and K. Eccleston, "ADAPTIVE DIGITAL PREDISTORTION LINEARSATION FOR RF POWER AMPLIFIERS", Electronics New Zealand Conference 2006

[9] F. Gregorio, and T. Laakso, "THE PERFORMANCE OF OFDM-CDMA SYSTEMS WITH POWER AMPLIFIER NON-LINEARITIES," Proceeding of the 2005 Finnish Signal Processing Symposium-FINSIG'05, Finland 2005

[10] A. Barbieri, D. Fertonani, and G. Colavolpe, " Spectrally-Efficient Continuous Phase Modulations", IEEE TRANSACTIONS ON WIRELESS COMMUNICATIONS, VOL. 8, NO. 3, MARCH 2009.

[11] A. Perotti, P. Remlein, and S. Benedetto, "Adaptive Coded Continuous-Phase Modulations for frequency-Division Multiuser Systems", ADVANCES IN ELECTRONICS AND TELECOMMUNICATIONS, VOL. 1, NO. 1, APRIL 2010.

[12] P. Varahram, S. Mohammady, M. Hamidon, R. Sidek, and S. Khatun,"DIGITAL PREDISTORTION TECHNIQUE FOR COMPENSATING MEMORY EFFECTS OF POWER AMPLIFIERS IN WIDEBAND APPLICATIONS", Journal of ELECTRICAL ENGINEERING, VOL. 60, NO. 3, PP. 129-135, 2009.

[13] C. Hua, A. Bo, and Z. Xiangmo,"Design of a BPNN Predistorter for Nonlinear HPA with Memory", Journal of Information & Computational Science Vol. 7, No. 4, pp. 863-868, 2010.